%% file: main.tex
\documentclass[conference]{IEEEtran}
\IEEEoverridecommandlockouts
\usepackage{tikz}
\usepackage[siunitx]{circuitikz}
\usetikzlibrary{decorations.pathreplacing,calligraphy}
\usepackage{cite}
\usepackage{amsmath,amssymb,amsfonts}
\usepackage{algorithmic}
\usepackage{graphicx}
\usepackage{textcomp}
\usepackage{xcolor}
\usepackage{bm}
\usepackage{enumitem}
\usepackage{subfiles}
\usepackage{float}
\usepackage{amsthm}
\usepackage{hyperref}
\usepackage{oplotsymbl}
\usepackage{amsmath}
\usepackage[font=footnotesize]{caption}
\usepackage[font=footnotesize]{subcaption}
\usepackage{subcaption}

\DeclareMathOperator*{\argmax}{argmax}
\newcommand{\norm}[1]{\left\lVert#1\right\rVert_\textrm{F}}

\newcommand*\xbar[1]{%
  \hbox{%
    \vbox{%
      \hrule height 0.5pt 
      \kern0.2ex
      \hbox{%
        \kern-0.1em
        \ensuremath{#1}%
        \kern-0.1em
      }%
    }%
  }%
}

\def\BibTeX{{\rm B\kern-.05em{\sc i\kern-.025em b}\kern-.08em
    T\kern-.1667em\lower.7ex\hbox{E}\kern-.125emX}}

\usepackage{tikz}

\begin{document}
\title{Physically Consistent Models for Intelligent Reflective Surface-assisted Communications under Mutual Coupling and Element Size Constraint
\thanks{This work was supported by the Discovery Grants Program of the Natural Sciences and Engineering Research Council of Canada (NSERC) and  Futurewei Technologies.}
}

\author{\IEEEauthorblockN{Mohamed Akrout$^1$, Faouzi Bellili$^1$, Amine Mezghani$^1$, Josef A. Nossek$^2$}
\IEEEauthorblockA{$^1$\textit{ECE Department}, \textit{University of Manitoba}, Winnipeg, Canada, \\ $^2$\textit{ECE Department}, \textit{Technische Universität München}, Munich, Germany \\
akroutm@myumanitoba.ca, \{Faouzi.Bellili,Amine.Mezghani\}@umanitoba.ca, josef.a.nossek@tum.de}
}

\maketitle

\begin{abstract}
We investigate the benefits of mutual coupling effects between the passive elements of intelligent reconfigurable surfaces (IRSs) on maximizing the achievable rate of downlink Internet-of-Things (IoT) networks. In this paper, we present an electromagnetic (EM) coupling model for IRSs whose elements are connected minimum scattering antennas (i.e., dipoles). Using Chu’s theory, we incorporate the finite antenna size constraint on each element of the IRS to obtain the IRS mutual impedance matrix. By maximizing the IRS phase shiters using the gradient ascent procedure, our numerical results show that mutual coupling is indeed crucial to avoid the achievable rate degradation when the spacing between IRS elements is down to a fraction of the wavelength.
\end{abstract}


\section{Introduction}
\subsection{Background and related work}
\begin{figure}[h!]
    \centering
    \includegraphics[scale=0.28]{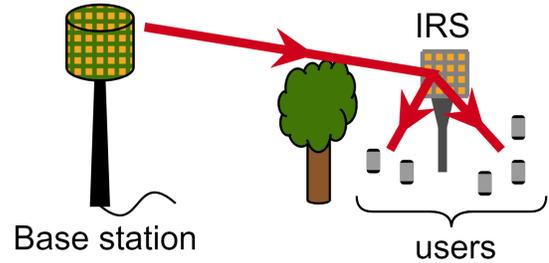}
    \caption{IRS-assisted downlink communication closing the coverage gap.}
    \label{fig:IRS-coverage}
    \vspace{-0.3cm}
\end{figure}

Both academia and industry are currently searching for future physical layer backbone technologies that can play pivotal roles in the design of beyond 5G (B5G) systems. Reconfigurable intelligent surfaces (IRSs) are regarded as one of these promising options to enable higher network capacity and peak data rates for next-generation wireless networks \cite{gong2020toward,di2020reconfigurable}. This is due to their unique capability of manipulating electromagnetic (EM) waves (e.g., steering, backscattering, and absorption) in a software-defined manner. Therefore, by favorably altering the propagation (e.g., fading, shadowing) of the radio waves, IRSs construct a strong channel between the transmitter and the receiver, thereby mitigating the coverage issue of future networks. A good example of this is when there are obstructions between the base station and the users as depicted in Fig.~\ref{fig:IRS-coverage}. IRSs are able to bend/deflect RF beams in 3D, thereby avoiding the obstructions
along the appropriate trajectories. Specifically, IRSs consist of planar arrays of passive scattering elements (e.g., small antenna), each of which is connected to a tunable chip to change its load impedance (e.g., PIN diode, varactor), thereby changing the EM properties of the impinging waves (e.g., the phase shift) \cite{di2020smart}.

The analysis of IRS-assisted wireless systems communication requires practical and accurate yet tractable models to properly account for the electromagnetic (EM) properties of the IRS elements. These models must also be consistent with two important physical considerations of the IRS elements: $i)$ the IRS element size constraint, and $ii)$ the mutual coupling between each pair of IRS elements. This research direction is still under an open investigation.

Little effort has been devoted to investigate the effect of mutual coupling (MC) on the performance of IRS-assisted wireless communication. Mutual coupling has been shown to be beneficial on the behavior of the eigen values of the spacial correlation of IRSs \cite{sun2022characteristics}. In \cite{gradoni2021end} and \cite{di2022modeling}, the authors proposed a scattering model that accounts for the mutual coupling between
the IRS elements. However, their model is not suitable for broadband communication where only lowest modes must be excited \cite{hansen2011small}. The same coupling model used in \cite{gradoni2021end} has been adopted in \cite{qian2021mutual} to maximize the intensity of the received voltage. There, by optimizing the IRS phase shifters, it has been showed that the received power can increase when MC is taken into account. While these physically consistent models proposed in the literature have considered the MC between the IRS elements, none of them accounts for the size constraint on each IRS element as a physical limitation. Note that the common constraint on the IRS dimension is different from the one imposed on each IRS element size as the latter restricts the radiation mode of the IRS elements \cite{hansen2011small}.

The form factor and aesthetic aspects of future
wireless equipments are indeed critical to keep wireless radiations and deployments compliant with those already imposed by regulations, while being acceptable by the general public. Moreover, the footprint of today’s wireless technology is dominated by the antenna size, which cannot be miniaturized beyond the Chu limit \cite{chu1948physical}. From this reason, the actual performance  of wireless communication systems can only be gauged if the antenna size is also considered as an integral part of their analysis, and IRSs are not the exception.

During the recent years, new antenna designs have emerged to enable multipurpose IRSs that can favorably alter the wave propagation. These antennas are ultrathin surfaces with subwavelength (i.e., microscopic) structures, also known as metasurfaces with 2D applied/induced magnetic and electric currents along the surface \cite{di2020smart}. In this work, we study IRSs composed of electric currents, namely, electrically small dipoles separated by a subwavelength spacing. This model of IRS elements leads to MC expressions that agrees with published results obtained by applying approximation techniques to structurally specified antennas \cite{wasylkiwskyj1970theory, borgiotti1968novel}.

\subsection{Contributions}
We incorporate both mutual coupling and the IRS element size constraint using Chu's theory \cite{chu1948physical}. We do so by capitalizing on the antenna size-aware mutual coupling model in \cite{akrout2022achievable}. The latter is used to obtain the mutual impedance matrix of the IRS, $\bm{Z}_S$. This matrix is used as part of the equivalent circuit model for radio communications to assess the achievable rate of the overall wireless IRS-assisted communication system. We also optimize the phase shifters to maximize the achievable rate using the gradient ascent procedure. Our simulation results show that mutual coupling is indeed beneficial to maintain the reflective performance of dense IRSs, i.e., with a subwavelength spacing between its elements.

\subsection{Paper organization and notations}
We structure the rest of this paper as follows. In Section \ref{sec:siso-system}, we present the circuit model for IRS-assisted MIMO communications and review its input-output relationship already derived in \cite{gradoni2021end}. We also present the coupling model between any antenna pair recently introduced in \cite{akrout2022achievable} which accounts for both mutual coupling and element size constraint based on Chu's theory. In Section \ref{sec:phase-shifter-opt}, we maximize the achievable rate of the overall IRS-assisted system using the gradient ascent algorithm after deriving the gradient in closed form. Simulation results are presented in Section \ref{sec:results} and we conclude the paper in Section \ref{sec:conclusion}.

We also mention the common notations used in this paper. Given any complex number $z$,  $\Re\{z\}$ and $\Im\{z\}$ return its real part and the imaginary part, respectively. Given any matrix $\bm{A}$, $\bm{A}^{\textsf{T}}$ and $\bm{A}^{*}$, and $\bm{A}^{\textsf{H}}$ refer to its transpose, conjugate, and hermitian, and $\text{diag}(\bm{A})$ returns a diagonal matrix by setting the off-diagonal elements of $\bm{A}$ to zero. Throughout the paper, $c$ denotes the speed of light in vacuum (i.e., $c \approx 3\times10^8$), $\lambda$ is the wavelength, and $k_b = 1.38 \times 10^{-23}\, \mathrm{m}^{2}\, \mathrm{kg} \,\mathrm{s}^{-2}\, \mathrm{K}^{-1}$ is the Boltzmann constant. Moreover, $\mu = 1.25\times 10^{-6} \,\mathrm{m}\, \mathrm{kg} \,\mathrm{s}^{-2}\, \mathrm{A}^{-2}$ and $\epsilon = 8.85\times 10^{-12} \,\mathrm{m}^{-3}\, \mathrm{kg}^{-1} \,\mathrm{s}^{4}\, \mathrm{A}^{2}$ are the permeability and permittivity of vacuum, respectively. Finally, $k_0=\omega\sqrt{\epsilon \, \mu}=\frac{2\pi}{\lambda}$ is the wave number in free space, respectively.

\section{Mutual coupling model for IRS}\label{subsec:mutual-coupling-IRS}
The IRS mutual coupling model used in \cite{gradoni2021end} relies on two assumptions: $i)$ the minimum scattering IRS elements and $ii)$ the thin cylindrical wire regime of the antenna elements made from perfectly conducting materials. In this section, we present a new coupling model recently introduced in \cite{akrout2022achievable} that relies on the minimum scattering assumption only and does not assume any specific antenna type. By not restricting the IRS elements to have a specific aperture form, this coupling model allows the information-theoretic assessment of the impact of the IRS elements on the overall performance of IRS-assisted MIMO communication system.

\subsection{Chu's CMS antenna model with size constraint}
\begin{figure}[h!]
\centering
\begin{circuitikz}[american voltages, american currents, scale=0.7, every node/.style={transform shape}]
\draw (0,0) node[anchor=east]{}
 to[short, o-*] (3,0);
 \draw (3,2) to[L, label=\mbox{\small{$L=\frac{a \,R}{c}$}}, *-*] (3,0);
 \draw (3,0) -- (5,0);
 \draw (5,2) to[/tikz/circuitikz/bipoles/length=30pt, R, l=\mbox{\small{$R$}}, -] (5,0);
 \draw (3,2) -- (5,2);
 \draw (0,2) node[anchor=east]{}
  to[C, i>_=\mbox{\small{$I_1(f)$}}, label=\mbox{\small{$C=\frac{a}{cR}$}}, o-*] (3,2);
  \draw[-latex] (1,0.5) -- node[above=0.05mm] {$Z_\text{Chu}(f)$} (2, 0.5);
  \draw (0,2) to [open,v=\small{$V_1(f)$}] (0,0);
\end{circuitikz}
\caption{Equivalent circuit of the TM$_1$ radiation mode of a Chu's antenna.}
\label{fig:tm1}
\end{figure}
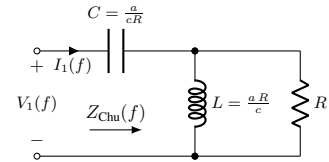

The seminal work of Chu \cite{chu1948physical} laid the foundations for the equivalent circuit models of any antenna whose structure can be embedded inside a spherical volume of a given radius $a$. More specifically, Chu derived an equivalent circuit network for each $n^{\text{th}}$ spherical $\text{TM}_{n}$ radiation mode of a given electromagnetic field in free space. For broadband communication applications, it suffices to consider antennas having the first radiation mode only (i.e., $n = 1$) since they have $i)$ the lowest Q-factor or equivalently the broadest bandwidth, and $ii)$ a gain of $3/2$ in the equatorial plane \cite[chapter 6]{harrington1961pp}. These are antennas with closed-form mutual impedances and are called canonical \textit{minimum scattering} (CMS) antennas \cite{kahn1965minimum}. Their equivalent ladder circuit network for the TM$_1$ mode is illustrated in Fig.~\ref{fig:tm1}.
\vspace{-0.3cm}

\noindent Using basic circuit analysis, one can write the self/input impedance $Z_\text{Chu}$ as:
\begin{equation}\label{eq:Z1-chu}
\begin{aligned}[b]
    Z_\text{Chu}(f) = \frac{V_1(f)}{I_1(f)}&= \underbrace{\frac{1}{\textrm{j}2\pi f\frac{a}{cR}}}_{Z_C(f)} + \underbrace{\frac{1}{\frac{1}{\textrm{j}2\pi f\frac{aR}{c}}+\frac{1}{R}}}_{Z_{\rm L}(f) \,\parallel\, Z_R(f)}\\
    &=\frac{c^2R + \textrm{j}2\pi fcaR - (2\pi fa)^2R}{\textrm{j}2\pi f ca - (2 \pi fa)^2}\,\, [\Omega].
\end{aligned}
\end{equation}

\subsection{Mutual coupling model between two Chu's CMS antennas}
\begin{figure}[h!]
    \centering
    \includegraphics[scale=0.78]{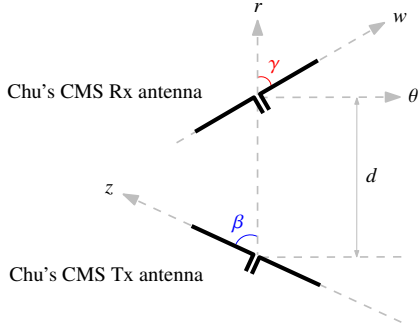}
    \caption{Two Chu's CMS antennas in the same plane, arbitrarily oriented in free space, and separated by a distance $d$ [m].}
    \label{fig:two-hertz-antennas}
\end{figure}

When the mutual coupling between antenna elements cannot be ignored due to the limited space, the mutual impedance is a practical measure of such proximity effects \cite{schelkunoff1952antennas}. For CMS antennas, the mutual impedance can be calculated analytically using the ``induced EMF'' method \cite[chapter 25]{orfanidis2002electromagnetic}. Based on a radiated power equivalence between Hertz dipoles and Chu's CMS antennas, it has been recently shown that the mutual impedance between two Chu's CMS antennas prescribed within non-overlapping spheres with self-impedances $Z_{\textrm{T}}(f)$ and  $Z_{\text{R}}(f)$ and separated by a distance $d$ are given by \cite{akrout2022achievable}:

\begin{equation}\label{eq:mutual-coupling-SISO}
\fontsize{9.5}{9.5}
    \begin{aligned}[b]
        &Z_{\text{TR}} = Z_{\text{RT}} =-3\,\sqrt{\Re\big[Z_{\text{T}}\big]\,\Re\big[Z_\text{R}\big]} \Bigg[\frac{1}{2}\,\sin(\beta)\,\sin(\gamma)\\
        &\hspace{1.9cm}\times\bigg(\frac{1}{\textrm{j}k_0d} + \frac{1}{(\textrm{j}k_0d)^2} +\frac{1}{(\textrm{j}k_0d)^3}\bigg)+\cos(\gamma)\,\cos(\beta)\\
        &\hspace{1.9cm}\times\bigg(\frac{1}{(\textrm{j}k_0d)^2} + \frac{1}{(\textrm{j}k_0d)^3}\bigg)\Bigg]\,e^{-\textrm{j}k_0d} ~\triangleq~ \mathcal{MC}(d,\beta,\gamma),
    \end{aligned}
\end{equation}
where the angles $\beta$ and $\gamma$ represent the rotation of the dipoles with respect to (w.r.t.) their connecting axis $r$ as depicted in Fig.~\ref{fig:two-hertz-antennas}. Throughout the paper, we refer to the mutual impedance expression in (\ref{eq:mutual-coupling-SISO}), parameterized by the parameter triplet ($d$, $\beta$, $\gamma$), as ``the mutual coupling model $\mathcal{MC}(d, \beta, \gamma)$'' between two Chu's CMS antennas.\vspace{-0.1cm}

\section{Physsically consistent models for IRS-assisted MIMO communications}\label{sec:siso-system}

\begin{figure}[H]
    \centering
    \input{./figs/multiport-mimo-IRS.tex}
    \vspace{-0.1cm}
    \caption{Equivalent circuit-theoretic model for IRS-assisted MIMO communication channels.}
    \label{fig:channel-equivalent-circuit-model}
    \vspace{-0.1cm}
\end{figure}
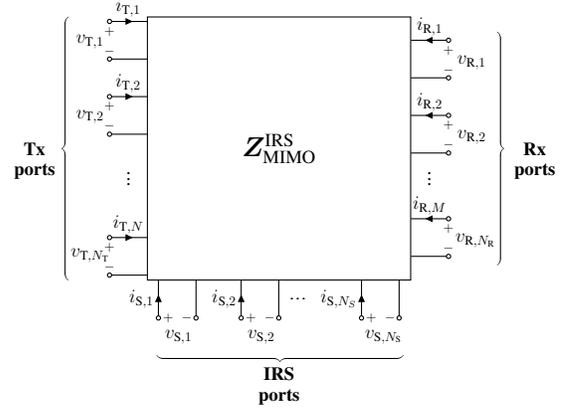

From circuit theory, transmitted/received signals are either voltages or currents that flow through the ports of the transmit/receive antennas. Finding the relationship between port variables at the transmitter(s) and receiver(s) is key to modelling both communication channels in a physically consistent way. As depicted in Fig. \ref{fig:channel-equivalent-circuit-model}, a IRS-assisted MIMO communication channel is modeled as a \textit{multiple-port network} with ($N_T+N_S+N_R$) ports representing $N_T$ transmit antennas, $N_S$ IRS elements, and $N_R$ receive antennas. Throughout this paper, we dedicate the letters and subscripts ``T'', ``S'', and ``R'' for the transmitter, the IRS, and the receiver, respectively.
\noindent We denote the voltage and current vectors of $\textrm{X}=\{T,S,R\}$ as $\bm{v}_{\textrm{X}}(f) = \big[ v_{\text{X},1}(f), v_{\text{X},2}(f), \dots, v_{\text{X},N_X}(f)\big]^\top$ and $\bm{i}_{\textrm{X}}(f) = \big[ i_{\text{X},1}(f), i_{\text{X},2}(f), \dots, i_{\text{X},N_X}(f)\big]^\top$. The relationship between the voltages ($\bm{v}_{\textrm{T}}$, $\bm{v}_{\textrm{S}}$, $\bm{v}_{\textrm{R}}$) and currents ($\bm{i}_{\textrm{T}}$, $\bm{i}_{\textrm{S}}$, $\bm{i}_{\textrm{R}}$) is given by Ohm's law:
\begin{equation}\label{eq:siso-V=ZI}
    \left[\begin{array}{l}
\bm{v}_\text{T}(f) \\
\bm{v}_\text{S}(f) \\
\bm{v}_\text{R}(f)
\end{array}\right]~=~\underbrace{\left[\begin{array}{ccc}
\bm{Z}_{\text{T}} & \bm{Z}_{\text{TS}}&\bm{Z}_{\text{TR}}\\
\bm{Z}_{\text{ST}} & \bm{Z}_{\text{S}} &\bm{Z}_{\text{SR}}\\
\bm{Z}_{\text{RT}} & \bm{Z}_{\text{RS}} &\bm{Z}_{\text{R}}
\end{array}\right]}_{\bm{Z}_{\text{MIMO}}^{\textrm{IRS}}(f)}\,\left[\begin{array}{l}
\bm{i}_\text{T}(f) \\
\bm{i}_\text{S}(f) \\
\bm{i}_\text{R}(f)
\end{array}\right].
\end{equation}
In (\ref{eq:siso-V=ZI}), the $N_{\textrm{T}}\times N_{\textrm{T}}$, $N_{\textrm{S}}\times N_{\textrm{S}}$ and $N_{\textrm{R}}\times N_{\textrm{R}}$ impedance matrices $\bm{Z}_{\text{T}}$, $\bm{Z}_{\text{S}}$ and $\bm{Z}_{\text{R}}$ stand for the impedances of the transmit array, the IRS, the receive array, respectively. Their diagonal elements correspond to the  transmit, IRS, and receive self-impedances (i.e., when all other antennas are absent) and their off-diagonal elements represent the mutual impedances between the antenna elements within the transmit array, the IRS, and the receive array, respectively.\vspace{-0.35cm}

\begin{figure*}[h!]
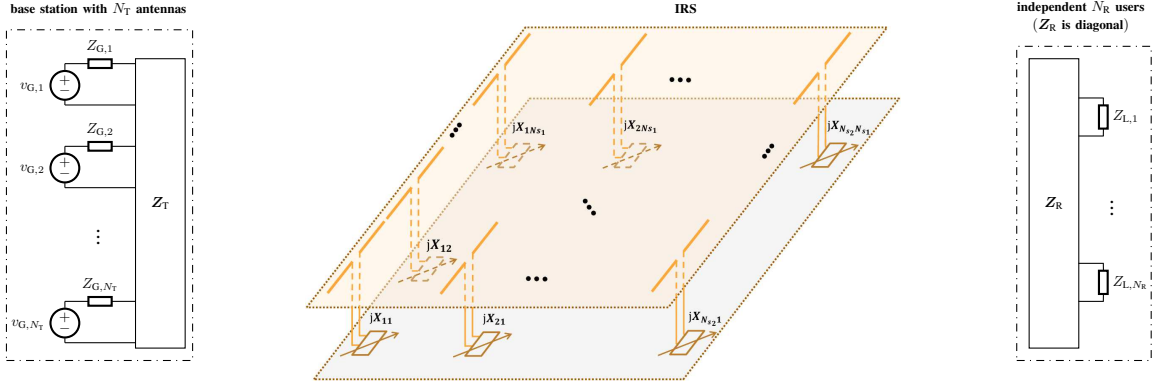

\centering
  \subfile{./figs/MIMO-IRS.tex}
\caption{Equivalent circuit model for IRS-assisted communication system: A base station with $N_{\textrm{T}}$ transmit antennas connecting with $N_{\textrm{R}}$ single-antenna users through a IRS composed of $N_S = N_{s_1} \times N_{s_2}$ connected dipoles. Every $ij$th dipole element is connected to a tunable impedance $Z_{\textrm{IRS},ij}=\textrm{j}\,X_{ij}$ controlling its phase shift. The dependence on the frequency argument ($f$) was dropped to lighten the notation in the figure.}
\label{fig:channel-circuit-models}
\rule{\textwidth}{0.4pt}
\vspace{-0.7cm}
\end{figure*}

\noindent For $X,X^{\prime}\in\{T,S,R\}$ and $\text{X}\neq\textrm{X}^{\prime}$, the matrices $\bm{Z}_{\text{X}\textrm{X}^{\prime}}$ are $N_{\textrm{X}} \times N_{\textrm{X}^{\prime}}$ symmetric matrices that gather the pairwise transimpedances between $X$ and $X^{\prime}$. Since antennas are reciprocal devices, it follows that $\bm{Z}_{\text{X}\textrm{X}^{\prime}} = \bm{Z}_{\text{X}^{\prime}\textrm{X}}^\top$ \cite{balanis}. Together, the aforementioned impedance and transimpedance matrices form the symmetric $(N_\text{T}+N_\text{S}+N_\text{R}) \times (N_\text{T}+N_\text{S}+N_\text{R})$ joint impedance matrix $\bm{Z}_{\text{MIMO}}^{\text{IRS}}$ of the IRS-assisted MIMO system. The knowledge of $\bm{Z}_{\text{MIMO}}^{\text{IRS}}$ fully characterizes the IRS-assisted MIMO communication system and enable its analysis with a compact multi-port matrix description.

\noindent To describe the circuit model of the IRS-assisted \textit{channel} in Fig. \ref{fig:channel-equivalent-circuit-model} as an integral part of a IRS-assisted MIMO communication \textit{system}, Fig. \ref{fig:channel-circuit-models} depicted the circuit models of the overall communication system composed of:
\begin{itemize}[leftmargin=*]
    \item the \textit{transmit side} as a base station equipped with $N_\textrm{T}$ non-ideal voltage generators $\{v_{\textrm{G},1},\dots,v_{\textrm{G},N_{\textrm{T}}}\}$ \big(i.e., with internal impedances $\{Z_{\textrm{G},1},\dots,Z_{\textrm{G},N_{\textrm{T}}}\}$\big). These voltage generators model the signal generation at each one of the $N_{\textrm{T}}$ transmit antenna ports.
    \item the \textit{IRS} which is consists of $N_{\textrm{S}}$ Chu's CMS antennas arranged over a 2D grid of size $N_{s_1}\times N_{s_2}$. Every $ij$th IRS element is connected to a tunable impedance $Z_{\textrm{IRS},ij}=\textrm{j}\,X_{ij}$ which acts as a phase shifter to the impinging wave. Each IRs element is embedded within inside a spherical volume of radius $a$. The inter-element spacing between all the Chu's spheres denoted $\delta$ is uniform within the IRS.
    \item the \textit{receiver side} composed of $N_{\textrm{R}}$ single-antenna users. Since their antennas are not in close proximity, the receive ports decompose into separate $N_{\textrm{R}}$ receive impedances due to the absence of MC, thereby resulting to a diagonal receive impedance matrix $\bm{Z}_{\textrm{R}}$ in Fig.~\ref{fig:channel-circuit-models}. \vspace{0.2cm}
\end{itemize}

\noindent For FF communications, signal attenuations in the FF region between the transmitter and the receiver, the transmitter and the IRS, and the IRS and the receiver are very large, i.e.:
\begin{subequations}\label{eq:unilateral-approx}
\begin{align}
     \norm{\bm{Z}_{\textrm{TR}}(f)} &= \norm{\bm{Z}_{\textrm{RT}}(f)} \ll \min\big(\norm{\bm{Z}_{\textrm{T}}(f)}, \norm{\bm{Z}_{\textrm{R}}(f)}\big),\\
     \norm{\bm{Z}_{\textrm{TS}}(f)} &= \norm{\bm{Z}_{\textrm{ST}}(f)} \ll \min\big(\norm{\bm{Z}_{\textrm{T}}(f)}, \norm{\bm{Z}_{\textrm{S}}(f)}\big),\\
     \norm{\bm{Z}_{\textrm{RS}}(f)} &= \norm{\bm{Z}_{\textrm{SR}}(f)} \ll \min\big(\norm{\bm{Z}_{\textrm{R}}(f)}, \norm{\bm{Z}_{\textrm{S}}(f)}\big).
\end{align}
\end{subequations}

\noindent This justifies the so-called ``unilateral approximation'' \cite{ivrlavc2010toward} which stipulates that the receive-transmit, IRS-transmit, and receive-IRS transimpedances $\bm{Z}_{\textrm{TR}}(f)$, $\bm{Z}_{\textrm{TS}}(f)$, and $\bm{Z}_{\textrm{SR}}(f)$, respectively can be ignored (i.e., $ \bm{Z}_{\textrm{TR}}(f) = \bm{Z}_{\textrm{TS}}(f)=\bm{Z}_{\textrm{SR}}(f) \approx \mathbf{0}$) in the FF region. This is because only the transmit antennas influence the electromagnetic properties of both the IRS and receive antennas.

\subsection{Transimpedance matrices}
The transimpedance matrices $\bm{Z}_\textrm{RT}(f)$, $\bm{Z}_\textrm{ST}(f)$, and $\bm{Z}_\textrm{RS}(f)$ describe the physically consistent channels that govern the electromagnetic influence of the transmitter on the receiver, the transmitter on the IRS, and the IRS on the receiver, respectively. Unlike conventional wireless channels, physically consistent channels account for the correlation effect due to the impedance matrices of the transmitter $X\in\{T,S\}$ and the receiver $Y\in\{S,R\}$ (with $\text{X}\neq\textrm{Y}$). Using basic circuit analysis and the Friis’ equation, it has been shown that the multi-path (MP) transimpedance matrix is given by \cite{akrout2022super}:
\begin{equation}\label{eq:transimpedance-multipath}
\fontsize{9.5}{9.5}
\begin{aligned}[b]
    \bm{Z}_{\mathrm{YX}}^{\textrm{MP}}(f) &= \frac{c\,\sqrt{G_\text{X}\,G_\text{Y}}}{2{\pi}fd^{\frac{\alpha}{2}}}~\textrm{diag}\left(\Re{\{\bm{Z_{\textrm{Y}}}(f)\}}\right)^{\frac{1}{2}}\\
    &\hspace{0.5cm}\times\sum\limits_{\ell=1}^{L}\bm{a}_\mathrm{Y}(\theta_{\mathrm{Y},\ell})\,\bm{a}_{\mathrm{X}}^{\mathsf{T}}(\theta_{\mathrm{T},\ell}) ~\textrm{diag}\left(\Re{\{\bm{Z_{\textrm{X}}}(f)\}}\right)^{\frac{1}{2}}\,e^{-j\phi}.
\end{aligned}
\end{equation}
In (\ref{eq:transimpedance-multipath}), $\alpha$ is the path-loss exponent and $d$ is the distance between the transmit antennas of $X$ and the receive antennas of $Y$ which have gains $G_{\textrm{X}}$ and $G_{\textrm{Y}}$ and sizes $a_{\textrm{X}}$ and $a_{\textrm{Y}}$, respectively. Moreover, $\theta_{\textrm{X},\ell}$ and $\theta_{\textrm{Y},\ell}$ are the angles of arrival and departure defined of the $\ell$th path w.r.t. the broadside axis of the transmit and receive arrays, respectively, and\vspace{-0.2cm}
\begin{subequations}
    \begin{align}
        \phi & = \pi - \textrm{arctan}\left(2\pi f a_{\textrm{X}}/c\right) - \textrm{arctan}\left(2\pi f a_{\textrm{Y}}/c\right),\nonumber\\
        \bm{a}_\mathrm{X}(\theta_{\mathrm{X},\ell})&=\bigg[1, e^{-\textrm{j}\frac{2\pi}{\lambda}\delta \cos(\theta_{\textrm{X},\ell})},\hdots, e^{-\textrm{j}\frac{2\pi}{\lambda}(N_{\textrm{X}}-1)\delta \cos(\theta_{\textrm{X},\ell})} \bigg]^\top,\nonumber\\
        \bm{a}_\mathrm{Y}(\theta_{\mathrm{Y}},\ell)&=\bigg[1, e^{-\textrm{j}\frac{2\pi}{\lambda}\delta \cos(\theta_{\textrm{Y}.\ell})},\hdots, e^{-\textrm{j}\frac{2\pi}{\lambda}(N_{\textrm{Y}}-1) \delta \cos(\theta_{\textrm{Y},\ell})} \bigg]\nonumber.
    \end{align}
\end{subequations}

\subsection{The input-output relationship of the channel model}
Using the unilateral approximation (\ref{eq:unilateral-approx}), it has been shown that the input-output relationship between the transmit voltage $\bm{v}_{\textrm{G}}$ and the load voltage at the receiver $\bm{v}_{\textrm{L}}$ is given by \cite{gradoni2021end}:
\begin{equation}\label{eq:input-output-relationship}
    \bm{v}_{\textrm{L}}(f) = \bm{H}(f)\,\bm{v}_{\textrm{G}}(f),
\end{equation}
where
\begin{equation}\label{eq:channel-model}
\begin{aligned}[b]
    \bm{H}(f) &= \left(\mathbf{I}_M + \bm{Z}_{\textrm{R}}(f)\,\bm{Z}_{\textrm{L}}(f)^{-1}\right)^{-1}\\
    &\hspace{0.2cm}\times\left(\bm{Z}_\textrm{RT}(f) - \bm{Z}_\textrm{RS} (\bm{Z}_\textrm{IRS}+\bm{Z}_\textrm{S})^{-1}  \bm{Z}_\textrm{ST}\right)\, (\bm{Z}_\textrm{G} + \bm{Z}_\textrm{T})^{-1}.
\end{aligned}
\end{equation}

\noindent In (\ref{eq:channel-model}), $\bm{Z}_\textrm{IRS}(f)$ is a diagonal matrix containing the IRS phase shifter impedance vector, $\bm{z}_{\textrm{IRS}}$, i.e.:
\begin{equation}\label{eq:phase-shiter-impedance}
\hspace{-0.3cm}\bm{Z}_\textrm{IRS}(f) = \textrm{diag}\left(\bm{z}_{\textrm{IRS}}\right)= \textrm{diag}\left([z_{\textrm{IRS},1}(f),\dots, z_{\textrm{IRS},N_S}(f)]\right),
\end{equation}

\noindent where $z_{\textrm{IRS},k}(f)$ is the $k$th IRS phase shifter impedance with $k\in\{1,\dots,N_S\}$. From now on, we will drop the frequency argument from voltages and impedances to lighten the notation. For ease of notation, we define the auxiliary quantities:\vspace{-0.1cm}
\begin{subequations}\label{eq:ABC-variables}
\begin{align}
    \bm{A} &~\triangleq~ \left(\mathbf{I}_{N_{\textrm{R}}} + \bm{Z}_{\textrm{R}}\,\bm{Z}_{\textrm{L}}^{-1}\right)^{-1}\bm{Z}_{\textrm{RT}}\,(\bm{Z}_\textrm{G} + \bm{Z}_\textrm{T})^{-1},\\
    \bm{B} &~\triangleq~ \left(\mathbf{I}_{N_{\textrm{R}}} + \bm{Z}_{\textrm{R}}\,\bm{Z}_{\textrm{L}}^{-1}\right)^{-1}\bm{Z}_{\textrm{RS}},\\
    \bm{C} &~\triangleq~ \bm{Z}_{\textrm{ST}}\,(\bm{Z}_\textrm{G} + \bm{Z}_\textrm{T})^{-1}.
\end{align}
\end{subequations}
Injecting back (\ref{eq:phase-shiter-impedance}) and (\ref{eq:ABC-variables}) into (\ref{eq:channel-model}), the channel $\bm{H}(f)$ becomes:
\begin{equation}\label{eq:channel-model-ABC}
    \bm{H}(\bm{z}_{\textrm{IRS}}) \triangleq \bm{A} - \bm{B}\,(\textrm{diag}\left(\bm{z}_{\textrm{IRS}}\right) + \bm{Z}_\text{S})^{-1}\,\bm{C}.
\end{equation}

\noindent Finally, we write the achievable rate of the IRS-assisted MIMO communication system as:
\begin{equation}\label{eq:capacity}
    C(\bm{z}_{\textrm{IRS}}) = \log_2\textrm{det}\Big(\mathbf{I} +\bm{H}(\bm{z}_{\textrm{IRS}})\,\bm{R}_{\textrm{T}}\,\bm{H}(\bm{z}_{\textrm{IRS}})^{\mathsf{H}} \Big),
\end{equation}
where $\bm{R}_{\textrm{T}} = \mathbf{I}_{N_\textrm{T}}/N_\textrm{T}$ is the transmit signal covariance matrix.
\section{IRS Phase Shifters Optimization}\label{sec:phase-shifter-opt}

In this section, we optimize the achievable rate of the IRS-assisted  MIMO communication system given in (\ref{eq:capacity}) under both the antenna size constraint and MC effects within the IRS. We do so by optimizing the phase shifts of the IRS which can be stated as the following non-convex optimization problem:
\begin{equation}\label{eq:capacity-formulation}
    \argmax_{\textrm{diag}(\bm{z}_{\textrm{IRS}})}~C(\bm{z}_{\textrm{IRS}}).
\end{equation}
We optimize (\ref{eq:capacity-formulation}) using the gradient ascent algorithm. We first write the gradient of $C(\bm{z}_{\textrm{IRS}})$ as (see Appendix \hyperref[appendix:gradient-derivation]{I}):

\begin{equation}\label{eq:grad-expression}
\begin{aligned}[b]
    \nabla_{\bm{z}_{\textrm{IRS}}}\,C(\bm{z}_{\textrm{IRS}}) &= \textrm{diag}\big(\bm{Q}(\bm{z}_{\textrm{IRS}})\,\bm{P}(\bm{z}_{\textrm{IRS}})\big)^*,
\end{aligned}
\end{equation}
\noindent where
\begin{subequations}\label{eq:grad-expression-variables}
\begin{align}
    \bm{P} &= -\log_2(e)\,\Big(\bm{E}\,\big(\bm{B}\,\bm{Q}\,\bm{E} + \bm{D}\big)^{-1}\,
    \bm{B}\Big)\bm{Q},\\
    \bm{Q} &= (\textrm{diag}(\bm{z}_{\textrm{IRS}})+\bm{Z}_S)^{-1},\\
    \bm{D} &= \mathbf{I} + \bm{A}\,\bm{R}_{\textrm{T}}\,\bm{A}^\mathsf{H} - \bm{A}\,\bm{R}_{\textrm{T}}\,\bm{C}^\mathsf{H}\,\bm{Q}\,\bm{B}^\mathsf{H},\\
    \bm{E} &= \bm{C}\,\bm{R}_{\textrm{T}}\,\big(\bm{B}\,\bm{Q}\,\bm{C}-
\bm{A}\big)^{\mathsf{H}}.
\end{align}
\end{subequations}

\noindent Using (\ref{eq:grad-expression}) and (\ref{eq:grad-expression-variables}), the gradient ascent update at time step $t$ is given by:
\begin{equation}\label{eq:gradient-ascent}
    \bm{z}_{\textrm{IRS}}^{t+1} = \bm{z}_{\textrm{IRS}}^{t} + \textrm{j}\,\alpha~ \Im\Big\{\textrm{diag}\Big(\bm{Q}(\bm{z}_{\textrm{IRS}}^t)\,\bm{P}(\bm{z}_{\textrm{IRS}}^t)\Big)^*\Big\},
\end{equation}
where $\alpha > 0$ is a step size parameter. Note that the projection of the gradient on the imaginary coordinate in (\ref{eq:gradient-ascent}) is due to the constraint that the IRS loads must be purely imaginary.
\section{Numerical results and discussions}\label{sec:results}
We numerically illustrate the impact of MC within the IRS on the achievable rate in (\ref{eq:capacity}) for the following three scenarios:
\begin{itemize}[leftmargin=*]
    \item[$1)$] \textit{MC model with random/optimized phase shifts} where all the entries of $\bm{Z}_S$ are computed using (\ref{eq:mutual-coupling-SISO}),
    \item[$2)$] \textit{decoupled model with random/optimized phase shifts} where $\bm{Z}_S$ from $1)$ is diagonal,
    \item[$3)$] \textit{MC model with mismatched phase shift optimization} where the optimized phase shifts obtained in $2)$ are used to assess the achievable rate using the matrix $\bm{Z}_S$ in $1)$.
\end{itemize}

\noindent Consider a base staion with $N_{\textrm{T}}=32$ antennas communicating to $N_{\textrm{R}}=5$ users via a reflective IRS node where no direct link is available, i.e., $\bm{Z}_{\textrm{RT}} =\mathbf{0}$. The carrier frequency is $30\,\textrm{GHz}$, the path-loss exponent is $2.5$, and the transmit power of the base station is $10\,\textrm{W}$. We assume a multipath propagation channel with 10 path components over a 40-meter distance between the base station and the RIS node and a line-of-sight propagation channel over 5-meter distance between the IRS node and the users. To solely focus on the mutual coupling effects within the IRS node, we consider a base station with decoupled and matched transmit antennas w.r.t. a reference impedance $R_0=50 \,[\Omega]$, i.e., $\bm{Z}_{\textrm{G}} = \bm{Z}_{\textrm{T}} = R_0\, \mathbf{I}_{N_\text{T}}$. We also consider users with matched antennas to their loads, i.e., $\bm{Z}_{\textrm{L}} = \bm{Z}_{\textrm{R}} = R_0\, \mathbf{I}_{N_\text{R}}$.

\noindent The IRS is composed of Chu's CMS antennas whose encompassing Chu's spheres have a radius $a_S$. To control the MC effects, we consider tightly coupled IRS elements (i.e., $\delta=2\,a_S$) and we increase or decrease the size of the IRS elements by varying $a_S\in [\lambda/20,\lambda/4]$. With a fixed IRS aperture of $4\lambda\times 4\lambda$, this is equivalent to varying the 2D grid of IRS elements from $8 \times 8$ to $27 \times 27$ elements.

\begin{figure}[h!]
    \centering
    \includegraphics[scale=0.3]{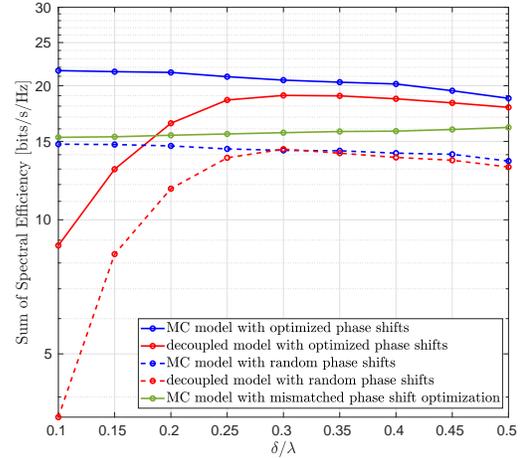}
    \vspace{-0.1cm}
    \caption{Performance of IRS-assisted 5-users downlink when the mutual coupling is ignored and taken into account.}
    \label{fig:spectral-efficiency-antenna-size}
    \vspace{-0.1cm}
\end{figure}

\noindent When the  mutual coupling is ignored, Fig.~\ref{fig:spectral-efficiency-antenna-size} shows a significant degradation of the spectral efficiency under both random and optimized phase shifters. If the mutual coupling is taking into account in the modelling of dense IRSs, the performance is flat due the spacial oversampling limit. In other words, just as time sampling at a rate exceeding the Nyquist criterion reduces quantization resolution requirements, IRSs with more tightly coupled elements exhibit better reflection performances than IRSs with decoupled elements.\vspace{-0.5cm}
 
\noindent Furthermore, when the optimized phase shifts using the decoupled model are employed to evaluate the achievable rate when MC is considered, we observe a lower spectral efficiency yet flat over the entire subwavelength spacing. This suggests that MC should be taken into account not only in the modelling of the IRS but also during the signal processing procedure of the IRS load tuning. From this perspective, this result confirms the need for a holistic design approach that examines the interplay between electromagnetics, signal processing, and communication theory to reach an optimal performance \cite{gabor1953communication}.

\section{Conclusion}\label{sec:conclusion}
 We presented an equivalent circuit model for IRS-assisted MIMO communications. By modelling the IRS passive elements as CMS Chu's antennas, we obtain both impedance and transimpedance matrices based on a new antenna coupling model that accounts for the IRS element size. We also optimized the achievable rate of the overall communication system at a single frequency using the gradient ascent procedure after deriving the close form expression of the gradient. We showed that our physically-consistent approach reveals beneficial MC effects for dense IRSs. This analysis can be extended to broadband communication scenarios where the achievable rate optimization has to account for larger bandwidth and impedance matching.

\begin{appendices}
\renewcommand{\thesectiondis}[2]{\roman{section}:}
\section{Derivation of the phase shift gradient}\label{appendix:gradient-derivation}
Consider the following non-convex optimization problem in (\ref{eq:capacity-formulation}).
where $\bm{H}$ is channel given in (\ref{eq:channel-model-ABC}). We first state the following lemma:\\

\noindent \textbf{Lemma~1.} \textit{Let} $\bm{Q}(\bm{z}_{\textrm{IRS}}) = (\textrm{diag}(\bm{z}_{\textrm{IRS}})+\bm{Z}_S)^{-1}$ \textit{where} $\bm{z}_{\textrm{IRS}}\in\mathbb{C}^{N_S}$ \textit{and} $\bm{Z}_S\in\mathbb{C}^{N_S\times N_S}$. \textit{We denote  the $k$th canonical basis vector in $\mathbb{R}^{N_S}$ as $\bm{e}_k=[0,\cdots,0,1,0,\cdots,0]^{\mathsf{T}}$, which has a single 1 in position $k$. Then, $\forall k\in\{1,\dots,N_S\}$, we have:}
\begin{equation*}\label{eq:lemma1}
    \frac{\partial\bm{Q}}{\partial z_{ \textrm{IRS},k}} = -(\textrm{diag}(\bm{z}_{\textrm{IRS}})+\bm{Z}_S)^{-1}\left(\bm{e}_k \,\bm{e}_k^\top\right)\,(\textrm{diag}(\bm{z}_{\textrm{IRS}})+\bm{Z}_s)^{-1}.
\end{equation*}
\begin{proof}
For a given matrix $\bm{G}(b)$ with $b\in\mathbb{C}$, we have the following identity: $\frac{\partial(\bm{G}^{-1})}{\partial b} = -\bm{G}^{-1}\frac{\partial\bm{G}}{\partial b}\,\bm{G}^{-1}$. Applying the latter to compute the derivative of $\bm{Q}$ w.r.t. the $k$th element of $\bm{z}_{\textrm{IRS}}$, $z_{ \textrm{IRS},k}$, yields:
\begin{equation*}
\begin{aligned}
    \frac{\partial\bm{Q}}{z_{ \textrm{IRS},k}} &= -\bm{Q} \,\frac{\partial(\textrm{diag}(\bm{z}_{\textrm{IRS}})+\bm{Z}_S)}{z_{ \textrm{IRS},k}}\,\bm{Q}\\
    &= -\bm{Q}\left(\bm{e}_k \, \bm{e}_k^\top\right)\bm{Q}.
\end{aligned}
\end{equation*}
\end{proof}

\noindent By defining $\bm{Q}(\bm{z}_{\textrm{IRS}}) \triangleq (\textrm{diag}(\bm{z}_{\textrm{IRS}})+\bm{Z}_S)^{-1}$ and using (\ref{eq:channel-model-ABC}), we start by rewriting $C(\bm{z}_{\textrm{IRS}})$ in (\ref{eq:capacity-formulation}) as:
\begin{equation}\label{eq:f-of-Z}
\begin{aligned}[b]
    C(\bm{z}_{\textrm{IRS}}) &= \log_2\textrm{det}\Big(\mathbf{I} +\bm{H}\,\bm{R}_{\textrm{T}}\,\bm{H}^{\mathsf{H}} \Big)\\
    &= \log_2\textrm{det}\Big(\mathbf{I} + \bm{A}\,\bm{R}_{\textrm{T}}\,\bm{A}^\mathsf{H} - \bm{A}\,\bm{R}_{\textrm{T}}\,\bm{C}^\mathsf{H}\,\bm{Q}(\bm{z}_{\textrm{IRS}})^\mathsf{H}\,\bm{B}^\mathsf{H}\\
    &\hspace{1.8cm}-\bm{B}\,\bm{Q}(\bm{z}_{\textrm{IRS}})\,\bm{C}\,\bm{R}_{\textrm{T}}\,\big(\bm{A}-\bm{B}\,\bm{Q}(\bm{z}_{\textrm{IRS}})\,\bm{C}\big)^{\mathsf{H}}\Big).
\end{aligned}
\end{equation}

\noindent Defining $\bm{D} \triangleq \mathbf{I} + \bm{A}\,\bm{R}_{\textrm{T}}\,\bm{A}^\mathsf{H} - \bm{A}\,\bm{R}_{\textrm{T}}\,\bm{C}^\mathsf{H}\,\bm{Q}(\bm{z}_{\textrm{IRS}})^\mathsf{H}\,\bm{B}^\mathsf{H}$ and $\bm{E} \triangleq \bm{C}\,\bm{R}_{\textrm{T}}\,\big(\bm{B}\,\bm{Q}(\bm{z}_{\textrm{IRS}})\,\bm{C}-
\bm{A}\big)^{\mathsf{H}}$, (\ref{eq:f-of-Z}) becomes:

\begin{equation}\label{eq:f-of-Z2}
\begin{aligned}[b]
    C(\bm{z}_{\textrm{IRS}})
    &= \log_2\textrm{det}\Big(\bm{D}+\bm{B}\,\bm{Q}(\bm{z}_{\textrm{IRS}})\,\bm{E}\Big).
\end{aligned}
\end{equation}

\noindent Using (\ref{eq:lemma1}), the differentiation chain rule, and the matrix derivative identity for arbitrary matrices $\bm{X}$, $\bm{S}$, $\bm{T}$ and $\bm{Y}$:
\begin{equation}
    \frac{\partial}{\partial \bm{X}} \log\textrm{det}\Big(\bm{S}\,\bm{X}\,\bm{T} + \bm{Y}\Big) = \Big(\bm{T}\,\big(\bm{S}\,\bm{X}\,\bm{T} + \bm{Y}\big)^{-1}\,
    \bm{S}\Big)^\top,\nonumber
\end{equation}

\noindent we take the Wirtinger derivative of (\ref{eq:f-of-Z2}) w.r.t. $\bm{z}_{\textrm{IRS}}$ to get the desired element-wise derivative expression:

\begin{equation}\label{eq:gradient-zk}
\begin{aligned}[b]
    \frac{\partial C(\bm{z}_{\textrm{IRS}})}{\partial z_{\textrm{IRS},k}} &= \operatorname{Tr}\Bigg(\left(\frac{\partial C(\bm{z}_{\textrm{IRS}})}{\partial \bm{Q}(\bm{z}_{\textrm{IRS}})}\right)^\top \cdot\, \frac{\partial \bm{Q}(\bm{z}_{\textrm{IRS}})}{\partial z_{\textrm{IRS},k}}\Bigg)\\
    &=\operatorname{Tr}\Bigg(-\log_2(e)\,\Big(\bm{E}\,\big(\bm{B}\,\bm{Q}\,\bm{E} + \bm{D}\big)^{-1}\,
    \bm{B}\Big)\\
    &\hspace{1.5cm}\times \Big(\bm{Q} \,\bm{e}_k\, \bm{e}_k^\top \, \bm{Q}\Big)\Bigg)\\
    &= \operatorname{Tr}\Big(\bm{P}\ \,\bm{e}_k\, \bm{e}_k^\top \, \bm{Q}\Big),
\end{aligned}
\end{equation}
\noindent  with $\bm{P} \triangleq \small{-\log_2(e)\,\big(\bm{E}\,\big(\bm{B}\,\bm{Q}\,\bm{E} + \bm{D}\big)^{-1}\,
    \bm{B}\big)\,\bm{Q}}$.

\noindent Finally, we write the full gradient w.r.t. the vector $\bm{z}_{\textrm{IRS}}$ by stacking all $k$-element derivatives in (\ref{eq:gradient-zk}) as:
\begin{equation}
\begin{aligned}[b]
    \nabla_{\bm{z}_{\textrm{IRS}}}\,C(\bm{z}_{\textrm{IRS}}) &= \left[\begin{array}{c}
\operatorname{Tr}\big(\bm{P}\,\bm{e}_1\, \bm{e}_1^\top\,\bm{Q} \big)\\
\vdots\\
\operatorname{Tr}\big(\bm{P}\,\bm{e}_{N_S}\, \bm{e}_{N_S}^\top\,\bm{Q}\big)
\end{array}\right]=\textrm{diag}\big(\bm{Q}\,\bm{P}\big).
\end{aligned}
\end{equation}

\end{appendices}

\bibliographystyle{IEEEtran}
\bibliography{IEEEabrv,references}

\end{document}

%% file: figs/multiport-mimo-IRS.tex
\begin{circuitikz}[american voltages, american currents, scale=0.5, transform shape]
\draw (0,0)
node[draw,minimum width=7cm,minimum height=7cm] (load) {\huge{$\bm{Z}_{\textrm{MIMO}}^{\textrm{IRS}}$}};
\draw ($(load.west)!0.75!(load.north west)+(0,0.75)$) coordinate (I1);
\draw ($(load.west)!0.75!(load.north west)+(0,-0.25)$) coordinate (I1back);
\draw ($(load.west)!0.75!(load.north west)+(0,-1.25)$) coordinate (I2);
\draw ($(load.west)!0.75!(load.north west)+(0,-2.25)$) coordinate (I2back);
\draw ($(load.west)!0.75!(load.north west)+(0,-5)$) coordinate (IM);
\draw ($(load.west)!0.75!(load.north west)+(0,-6)$) coordinate (IMback);
\draw ($(load.east)!0.75!(load.north east)+(0,0.75)$) coordinate (Ir1);
\draw ($(load.east)!0.75!(load.north east)+(0,-0.25)$) coordinate (Ir1back);
\draw ($(load.east)!0.75!(load.north east)+(0,-1.25)$) coordinate (Ir2);
\draw ($(load.east)!0.75!(load.north east)+(0,-2.25)$) coordinate (Ir2back);
\draw ($(load.east)!0.75!(load.north east)+(0,-5)$) coordinate (IrM);
\draw ($(load.east)!0.75!(load.north east)+(0,-6)$) coordinate (IrMback);
\draw ($(I1) + (-1,0)$) to[short,i>^={\Large$i_{\text{T},1}$},o-] (I1);
\draw ($(I1back) + (-1,0)$) to[short,o-] (I1back);
\draw ($(I2) + (-1,0)$) to[short,i>^={\Large$i_{\text{T},2}$},o-] (I2);
\draw ($(I2back) + (-1,0)$) to[short,o-] (I2back);
\draw ($(IM) + (-1,0)$) to[short,i>^={\Large$i_{\text{T},N}$},o-] (IM);
\draw ($(IMback) + (-1,0)$) to[short,o-] (IMback);
\draw ($(Ir1) + (1,-0.5)$) to[short,i>_={\Large$i_{\text{R},1}$},o-] ($(Ir1)+ (0,-0.5)$);
\draw ($(Ir1back) + (1,-0.5)$) to[short,o-] ($(Ir1back)+ (0,-0.5)$);
\draw ($(Ir2) + (1,-0.5)$) to[short,i>_={\Large$i_{\text{R},2}$},o-] ($(Ir2) + (0, -0.5)$);
\draw ($(Ir2back) + (1,-0.5)$) to[short,o-] ($(Ir2back) + (0, -0.5)$);
\draw ($(IrM) + (1,0.5)$) to[short,i>_={\Large$i_{\text{R},M}$},o-] ($(IrM) + (0,0.5)$);
\draw ($(IrMback) + (1,0.5)$) to[short,o-] ($(IrMback) + (0,0.5)$);
\draw ($(I1) + (0.3,-7.89)$) to[short,i>^={\Large$i_{\text{S},1}$},o-] ($(I1) + (0.3,-6.89)$);
\draw ($(I1) + (1.3,-7.89)$) to[short,o-] ($(I1) + (1.3,-6.89)$);

\draw ($(I1) + (2.5,-7.89)$) to[short,i>^={\Large$i_{\text{S},2}$},o-] ($(I1) + (2.5,-6.89)$);
\draw ($(I1) + (3.5,-7.89)$) to[short,o-] ($(I1) + (3.5,-6.89)$);

\draw ($(I1) + (5.7,-7.89)$) to[short,i>^={\Large$i_{\text{S},N_S}$},o-] ($(I1) + (5.7,-6.89)$);
\draw ($(I1) + (6.7,-7.89)$) to[short,o-] ($(I1) + (6.7,-6.89)$);

\draw ($(I1) + (-1,-0.15)$) to [open,v={\small }] ($(I1) + (-1,-0.9)$);
\draw ($(I2) + (-1,-0.15)$) to [open,v={\small }] ($(I2) + (-1,-0.9)$);
\draw ($(IM) + (-1,-0.15)$) to [open,v={\small }] ($(IM) + (-1,-0.9)$);
\draw ($(Ir1) + (1,-0.15-0.5)$) to [open,v={\small }] ($(Ir1) + (1,-0.9-0.5)$);
\draw ($(Ir2) + (1,-0.15-0.5)$) to [open,v={\small }] ($(Ir2) + (1,-0.9-0.5)$);
\draw ($(IrM) + (1,-0.15+0.5)$) to [open,v={\small }] ($(IrM) + (1,-0.9+0.5)$);

\node[] at (-1.5-2.5,-0.7) {{\textbf{$\vdots$}}};
\node[] at (1.5+2.5,-0.7) {{\textbf{$\vdots$}}};
\node[] at (-2.4-2.5,2.85) {{\Large$v_{\text{T},1}~~$}};
\node[] at (-2.4-2.5,0.85) {{\Large$v_{\text{T},2}~~$}};
\node[] at (-2.4-2.5,-2.85) {{\Large$v_{\text{T},N_\text{T}}~~$}};
\node[] at (2.4+2.6,2.85-0.55) {{\Large$~~v_{\text{R},1}$}};
\node[] at (2.4+2.6,0.85-0.55) {{\Large$~~v_{\text{R},2}$}};
\node[] at (2.4+2.7,-2.85+0.4) {{\Large$~~v_{\text{R},N_\text{R}}$}};

\node[] at (-2.95,-4.5) {{$+$}};
\node[] at (-2.45,-4.5) {{$-$}};
\node[] at (-2.55,-5) {{\Large$v_{\text{S},1}~~$}};

\node[] at (-0.75,-4.5) {{$+$}};
\node[] at (-0.25,-4.5) {{$-$}};
\node[] at (-0.35,-5) {{\Large$v_{\text{S},2}~~$}};

\node[] at (2.45,-4.5) {{$+$}};
\node[] at (2.95,-4.5) {{$-$}};
\node[] at (2.9,-5) {{\Large$v_{\text{S},N_\text{S}}~~$}};

\node[] at (0.5,-4) {{{$\mathbf{\hdots}$}}};
\node[] at (0.505,-4) {{{$\mathbf{\hdots}$}}};
\node[] at (0.495,-4) {{{$\mathbf{\hdots}$}}};

\draw [decorate, decoration = {calligraphic brace}] (5.8,3.1) --  (5.8,-3.1);
\node[] at (6.8,0) {\Large \textbf{Rx}};
\node[] at (6.8,-0.6) {\Large \textbf{ports}};

\draw [decorate, decoration = {calligraphic brace}] (-5.6,-3.5) --  (-5.6,3.5);
\node[] at (-6.5,0) {\Large \textbf{Tx}};
\node[] at (-6.5,-0.6) {\Large \textbf{ports}};

\draw [decorate, decoration = {calligraphic brace}] (3.3,-5.5) --  (-3.25,-5.5);
\node[] at (0,-6) {\Large \textbf{IRS}};
\node[] at (0,-6.6) {\Large \textbf{ports}};

\end{circuitikz}

%% file: figs/MIMO-IRS.tex
\begin{circuitikz}[american voltages, american currents, scale=0.55, transform shape]


\draw (0,0)
node[draw,minimum width=2cm,minimum height=7cm,opacity=0] (load) {}; 
\draw ($(load.west)!0.75!(load.north west)+(0,0.75)$) coordinate (I1);
\draw ($(load.west)!0.75!(load.north west)+(0,-0.25)$) coordinate (I1back);
\draw ($(load.west)!0.75!(load.north west)+(0,-1.25)$) coordinate (I2);
\draw ($(load.west)!0.75!(load.north west)+(0,-2.25)$) coordinate (I2back);
\draw ($(load.west)!0.75!(load.north west)+(0,-5)$) coordinate (IM);
\draw ($(load.west)!0.75!(load.north west)+(0,-6)$) coordinate (IMback);
\draw ($(load.east)!0.75!(load.north east)+(0,0.75)$) coordinate (Ir1);
\draw ($(load.east)!0.75!(load.north east)+(0,-0.25)$) coordinate (Ir1back);
\draw ($(load.east)!0.75!(load.north east)+(0,-1.25)$) coordinate (Ir2);
\draw ($(load.east)!0.75!(load.north east)+(0,-2.25)$) coordinate (Ir2back);
\draw ($(load.east)!0.75!(load.north east)+(0,-5)$) coordinate (IrM);
\draw ($(load.east)!0.75!(load.north east)+(0,-6)$) coordinate (IrMback);

\draw ($(I1) + (-4.5,-0.15)$) to[/tikz/circuitikz/bipoles/length=33pt, V, l_=$v_{\textrm{G},1}$] ($(I1) + (-4.5,-0.9)$);
\draw ($(I1back) + (-4.5,0)$) to[short,-] ($(I1back) + (-2.8,0)$);
\draw ($(I1back) + (-4.5,0)$) to ($(I1) + (-4.5,-0.9)$);
\draw ($(I1) + (-4.5,-0.15)$) to ($(I1) + (-4.5,0)$);
\draw ($(I1) + (-4.5,0)$)  to[european resistor, /tikz/circuitikz/bipoles/length=20pt,l=$Z_{\textrm{G},1}$,-] ($(I1)+ (-2.8,0)$);

\draw ($(I2) + (-4.5,-0.15)$) to[/tikz/circuitikz/bipoles/length=33pt, V, l_=$v_{\textrm{G},2}$] ($(I2) + (-4.5,-0.9)$);
\draw ($(I2back) + (-4.5,0)$) to[short,-] ($(I2back) + (-2.8,0)$);
\draw ($(I2back) + (-4.5,0)$) to ($(I2) + (-4.5,-0.9)$);
\draw ($(I2) + (-4.5,-0.15)$) to ($(I2) + (-4.5,0)$);
\draw ($(I2) + (-4.5,0)$)  to[european resistor, /tikz/circuitikz/bipoles/length=20pt,l=$Z_{\textrm{G},2}$,-] ($(I2)+ (-2.8,0)$);

\draw ($(IM) + (-4.5,-0.15)$) to[/tikz/circuitikz/bipoles/length=33pt, V, l_=$v_{\textrm{G},N_\textrm{T}}$] ($(IM) + (-4.5,-0.9)$);
\draw ($(IMback) + (-4.5,0)$) to[short,-] ($(IMback) + (-2.8,0)$);
\draw ($(IMback) + (-4.5,0)$) to ($(IM) + (-4.5,-0.9)$);
\draw ($(IM) + (-4.5,-0.15)$) to ($(IM) + (-4.5,0)$);
\draw ($(IM) + (-4.5,0)$)  to[european resistor, /tikz/circuitikz/bipoles/length=20pt,l=$Z_{\textrm{G},N_{\textrm{T}}}$,-] ($(IM)+ (-2.8,0)$);
\draw (-3.2,0)
node[draw,minimum width=1.2cm,minimum height=7cm] (load) {$\bm{Z}_{\textrm{T}}$};

\node[inner sep=0pt] (whitehead) at (7.5,0)
    {\includegraphics[scale=0.6]{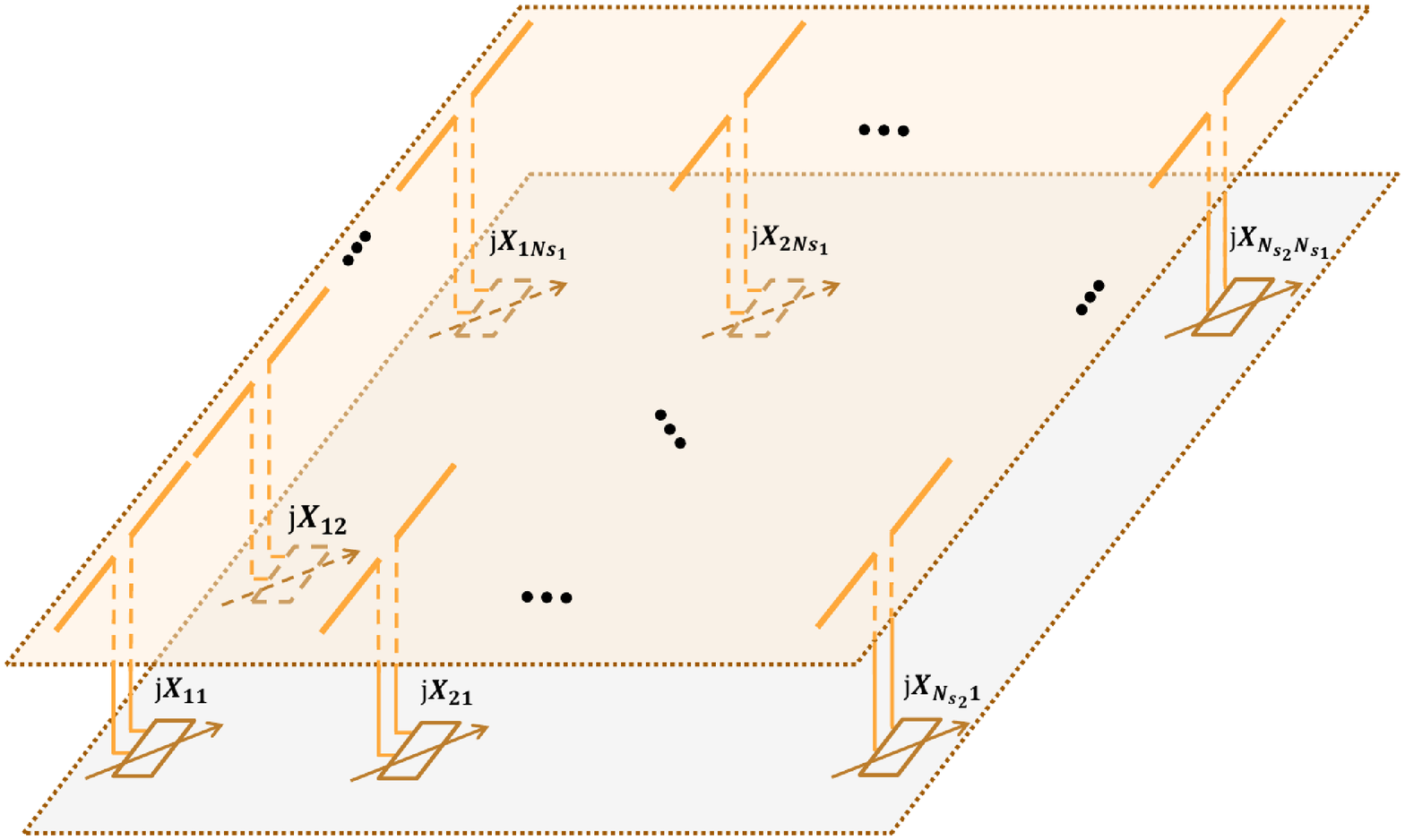}};

\draw ($(IMback) + (-4+24,0+5)$) to[short,-] ($(IMback) + (-4+24.61,0+5)$);
\draw ($(IMback) + (-4+24,0.92+5)$) to[short,-] ($(IMback) + (-4+24.61,0.92+5)$);

\draw ($(IM) + (-3.9+24.5,-0.08+5)$) to[european resistor, /tikz/circuitikz/bipoles/length=20pt, l^=$Z_{\textrm{L},1}$] ($(IMback) + (-3.9+24.5,0+5)$);

\draw ($(IMback) + (-4+24,0+1)$) to[short,-] ($(IMback) + (-4+24.61,0+1)$);
\draw ($(IMback) + (-4+24,0.92+1)$) to[short,-] ($(IMback) + (-4+24.61,0.92+1)$);

\draw ($(IM) + (-3.9+24.5,-0.08+1)$) to[european resistor, /tikz/circuitikz/bipoles/length=20pt, l^=$Z_{\textrm{L},N_{\textrm{R}}}$] ($(IMback) + (-3.9+24.5,0+1)$);
\draw (18.4,0)
node[draw,minimum width=1.2cm,minimum height=7cm] (load) {$\bm{Z}_{\textrm{R}}$};

\draw[dashdotted] (-6.92,-3.8) rectangle +(4.5,8);
\node[] at (-6.85+2.15,-3.8+8.5) {$\small{\textrm{\textbf{base station with~}} N_{\textrm{T}} \textrm{\textbf{~antennas}} }$};
\draw[dashdotted] (17.5,-3.8) rectangle +(3.25,7.6);
\node[] at (19.05,-3+7.75) {$\small{\textrm{\textbf{independent~}} N_{\textrm{R}} \textrm{\textbf{~users}} }$};
\node[] at (19.05,-3+7.3) {$\small{ (\bm{Z}_{\textrm{R}}\textrm{\textbf{~is diagonal}})}$};

\node[] at (9.5,4.7) {$\small{\textrm{\textbf{IRS}}}$};

\node[] at (-4.7,-0.7) {{\textbf{$\vdots$}}};
\node[] at (-4.7+24.5,0) {{\textbf{$\vdots$}}};

\end{circuitikz}
